\documentclass[12pt]{article}
\usepackage{natbib}
\usepackage{url}
\usepackage{lmodern}			
\usepackage[utf8]{inputenc}		
\usepackage{indentfirst}		
\usepackage{nomencl} 			
\usepackage{color}				
\usepackage{graphicx}			
\usepackage{microtype} 			
\usepackage{amsthm,amsmath,bbm} 
\usepackage{tabu} 
\usepackage{multirow}
\usepackage{rotating}
\usepackage{pdflscape}
\usepackage{authblk}
\usepackage{longtable}
\usepackage[left=2cm, right=2.5cm, top=2cm]{geometry}
\usepackage{booktabs}
\usepackage{float}
\usepackage{hyperref}
\hypersetup{
	colorlinks=true,
	linkcolor=blue,
	filecolor=blue,      
	urlcolor=blue,
	citecolor=blue
}

\usepackage{setspace}
\usepackage{titling}
\title{Underreporting of Intimate Partner Violence in Brazil}

\author{Diego de Maria André\thanks{Professor at the Graduate Program in Economics (PPECO), Federal University of Rio Grande do Norte (UFRN) and Intelligence Analyst at CENSIPAM/Ministry of Defense. E-mail:diego.andre@ufrn.br. ORCID: 0000-0003-3142-8336. (Author for correspondence)}  and José Raimundo Carvalho\thanks{Full Professor at the Graduate Program in Economics, CAEN/UFC and Intelligence Analyst at CENSIPAM/Ministry of Defense. E-mail: josecarv@ufc.br. ORCID: 0000-0001-5774-5925.}}
\date{April, 2025}

\begin{document}
	
\maketitle

\begin{abstract}
\noindent According to \citet{Organization2013}, in general 30\% of all women worldwide who have been in a relationship have experienced physical and/or sexual violence by their intimate partner. However, only a small percentage of intimate partner violence (IPV) victims report it to the police. This phenomenon of under-reporting is known as ``dark figure''. This paper aims to investigate the factors associated with the reporting decision of IPV victims to the police in Brazil using the third wave of the \textit{Pesquisa de Condi\c{c}\~{o}es Socioecon\^{o}micas e Viol\^{e}ncia Dom\'{e}stica e Familiar contra a Mulher ($PCSVDF^{Mulher}$)}. Using a bivariate probit regression model with sample selection, we found that older white women, those who do not tolerate domestic violence, and women who have experienced physical violence are more likely to report IPV to the police. In contrast, married women, those with partners who abuse alcohol and those who witnessed or knew that their mothers had experienced IPV, are less likely to report it to law enforcement.

\vspace{0.5cm}
\end{abstract}

\newpage

\section{Introduction}\label{intro}

Intimate partner violence (IPV) is defined as any act of physical violence, sexual violence, stalking, and psychological aggression, including coercive tactics, by a current or former intimate partner, such as a spouse, boyfriend/girlfriend, dating partner, or ongoing sexual partner \citep{Breiding2015}. According to \citet{Organization2013}, overall, 30\% of all women worldwide who have been in a relationship have experienced physical and/or sexual violence by their intimate partner \citep{Organization2013}. Facing these statistics, the World Health Organization recognized violence against women as a ``global health problem of epidemic proportions''\citep{Organization2013}.

Despite these alarming numbers, only a small fraction of women victims of IPV seek help and denounce their aggressor. Only between 2. 5\% and 15\% of IPV victims report having suffered this type of violence \citep{Gracia2004}. This phenomenon, known as the ``dark figure'', limits the capacity of public authorities to address the problem, since the real impacts of IPV on society are not known \citep{skogan_contacts_1994}. Therefore, understanding the phenomenon of underreporting and what leads a victim of IPV to report or not report the case to the authorities is fundamental to implementing effective and efficient public policies to prevent and combat IPV.

Some studies, such as \citet{skogan_contacts_1994} and \citet{macdonald_revisiting_2001}, seek to understand the factors that lead a victim of a crime to report or not a crime to the police. In general, they assume that the victims have rational behavior, i.e. they will report the crime if the reporting incentives are greater than the cost. So, if the perceived outcome of the report is negligible, one could expect that individuals would not report a crime \citep{macdonald_revisiting_2001}. In the context of IPV, there are several well-documented barriers in the literature to victims reporting domestic violence to the police, such as fear of not being believed by the police, the desire to protect the offender, the fear of removal of children and/or dissolution of the relationship, the economic dependence on the perpetrator, privacy concerns, fear of exposing their own illegal activities and fear of reprisal and the escalation of the violence \citep{Voce2018}. However, self-protection, history of victimization, details of the incident, such as drunken offender, severity of violence, and witnesses of IPV as children, are factors that increase the probability of reporting violence to the police \citep{felson_reasons_2002, Voce2018}.

There are several studies around the world, such as \citet{felson_reasons_2002}, \citet{akers_police_2009}, \citet{podana_reporting_2010}, \citet{kim_factors_2022}, \citet{Basterretxea2024}, among others, that seek to explain the woman's decision to report IPV. However, to our knowledge, there are no quantitative studies in Brazil that seek to explain the woman's decision to report IPV. So, to fill this gap and shed light on this important issue, this paper uses data from the third wave of the $PCSVDF^{Mulher}$\footnote{\textit{Pesquisa de Condi\c{c}\~{o}es Socioecon\^{o}micas e Viol\^{e}ncia Dom\'{e}stica e Familiar contra a Mulher ($PCSVDF^{Mulher}$)}}, the most extensive survey on domestic violence against women carried out in Latin America. Through face-to-face interviews, $PCSVDF^{Mulher}$ collected information, among others, on victimization and reporting decisions of 8447 women in the cities of Fortaleza/CE, Recife/PE, Salvador/BA, Goiânia/GO, Porto Alegre/RS, São Paulo/SP and Belém/PA between August and December 2019.

To understand the phenomenon of the ``dark figure'' of IPV in Brazil, we followed two steps: The first was to analyze the responses given by women who were victims of domestic violence about the reasons that led them to report or not the violence they suffered to the police. After this first moment, following \citet{macdonald_revisiting_2001}, we estimate a bivariate probit regression model with sample selection to identify the factors that influence the reporting decision.

In addition to this introduction, the remainder of this paper is structured as follows. Section \ref{background} presents a brief review of the theories that seek to explain the woman's decision to report IPV to the police and some empirical studies on it. Section \ref{methods} introduces the data set and the measures used in our estimates. We also present the model used to estimate the factors that influence the decision to report IPV. Section \ref{results} presents the results and Section \ref{discussion} discusses the results obtained by the estimation of the model presented in the previous section. We conclude with a brief conclusion.

\section{Background}\label{background}

Under-reporting of IPV occurs when victims do not report an incident of violence to the police or do not seek help \citep{lohani2021}. Unfortunately, the vast majority of IPV cases go unreported, and only a small fraction of the cases are reported to the police. \citet{Gracia2004} calls this phenomenon the ``iceberg'' of domestic violence. This large number of invisible IPV cases can cause great harm to society since the real extent of the problem is unknown, hindering the design of more effective and efficient public policies to combat and prevent the problem. Therefore, it is essential to understand the causes of under-reporting of IPV cases.

Some theories aim to explain why a woman who is a victim of IPV might choose to report or not report the abuse she has experienced. For example, \textit{Social Learning theory} suggests that aggression is learned by observation of the behavior of others, especially within the family \citep{Copp2019}. Thus, this theory suggests that victims of IPV have learned to accept violence as a norm and, therefore, are less likely to report violence to the police.

Another theory that tries to explain the victim's decision to report or not violence is the \textit{Secondary Victimization theory}. Secondary victimization is defined as the second victimization that occurs after the violent episode, and is the result of people's judgments regarding the victim and results in the absence of social support and negative judgments, and may be more painful and traumatic than the IPV itself, as it is perpetrated by those who should be the caregivers or protectors of the victim \citep{Oliveira2024}.

\citet{akers_police_2009} summarize the feminist perspectives in the decision to report or not IPV. The \textit{feminist theory} holds that women are considered ``culturally legitimate victims''. In marriage and marriage-like relationships, the partner controls the couple's resources, preventing the woman from having free access to and control over their properties. This control may explain why some IPV victims seek help, and some do not. Another point of view from the feminist perspective is that the perpetuation of men's power over women may make police officers and court officials reluctant to believe women who report incidents of IPV and hesitant to enforce existing arrest laws, which may discourage women from seeking help once they may feel that she will not be believed and their abusers will not be punished. Finally, marital dependency, which can increase due to the lack of educational and financial resources and the presence of young children, can reduce the possibility of the woman leaving a violent relationship and may also decrease the probability of reporting the experience of violence.

In the \textit{rational choice framework}, the victim's decision to report or not an IPV is influenced more by a sort of personal cost-benefit analysis \citep{gottfredson_victims_1988}. Victims are more likely to report an IPV to the police when incentives are high and costs are low \citep{felson_reasons_2002}. We can cite incentives to stop an ongoing attack, deter future attacks, desire for retribution or justice, protect other people, such as children, and the severity of the incident. The more serious the incident, with the use of weapons or serious injuries, the greater the likelihood that the victim will report the case to the police. However, the victim's cost can include embarrassment and protection of status, the desire to protect the offender from prison, and fear of reprisals \citep{felson_reasons_2002}. 

Having discussed the main theories explaining a woman's decision to report or not report the IPV she experienced to the police, we will now review empirical studies that have investigated the underreporting of IPV cases.

Using data from the National Crime Victimization Survey (USA), \citet{felson_reasons_2002} analyzed the decision of reporters and non-reporters of domestic violence using an approach based on incentives and costs. \citet{felson_reasons_2002} found that privacy concerns, fear of reprisal, and desire to protect offenders are factors that prevent victims of domestic violence from calling the police. At the same time, self-protection and perceived domestic assaults as more serious are factors that encourage reporting. \citet{felson_reasons_2002} also found that victims are likely to view domestic assaults as more threatening and serious because they tend to occur at home.

\citet{akers_police_2009} analyzed how police notification decisions are influenced by the demographic characteristics of victims and incident-specific factors. Using data from the 1999 Canadian General Social Survey, Victimization (CAN), \citet{akers_police_2009} found that victims who live in a cohabiting relationship and with children are more willing to report IPV to the police, while income, education and employment status do not appear to influence reporting decisions. \citet{akers_police_2009} also found that, contrary to predictions, visible minority women more often call the police. Concerning the incident-specific factors, \citet{akers_police_2009} found that women who call the police are also likely to have experienced severe forms of violence, including threats with weapons, injury, and the destruction of their property.

In Europe, \citet{podana_reporting_2010} analyzed the reporting decisions of victims of domestic violence in the Czech Republic. For this, \citet{podana_reporting_2010} utilized the Czech part of the International Violence Against Women Survey. \citet{podana_reporting_2010} found that the reporting rate among the victims of IPV was extremely low, with only 8\% of the victims of IPV reporting the case to the authorities. Using a logistic regression model, \citet{podana_reporting_2010} found that features of the particular incident and factors related to the history of violence in the relationship are relevant to explain the decision to report. At the same time, \citet{podana_reporting_2010} also found that distrust of the police is an important factor in deciding not to report to the police.

In Asia, \citet{kim_factors_2022} examined the factors associated with the willingness to report IPV to police for both men and women using data from the 2013 Korean National Domestic Violence Survey. For both men and women, \citet{kim_factors_2022} found that the willingness to report IPV to the police was statistically significant when the participants were young, had a strong knowledge of IPV-related laws, had lower levels of acceptance of violence, had lower levels of values of conservative gender roles, and when the severity of violence was greater. However, having children and having experienced child abuse are factors that exclusively affected women's willingness to report IPV to the police.

In South America, \citet{Basterretxea2024} examined the effect of perceived reportability of IPV on women's decision to report IPV to the police. Using data from the Chilean National Institute of Youth (CHI), \citet{Basterretxea2024} showed that perceived reportability was a significant predictor of choosing the police as a source of help compared to other informal sources of help, such as family and friends, but has no effect on the choices between the police and other formal sources of help, such as psychologists and public services. \citet{Basterretxea2024} also found that participants who cohabit or have married married (thus assumed to cohabite) were more likely to seek the help of the police in the event of IPV than participants in less stable or more informal relationships, such as dating or short-term relationships.

After reviewing the main theories that explain why a woman who is a victim of domestic violence might decide to report the incident to the police, along with some empirical studies on the topic, in the next section, we will present the dataset and methodology used in this article.

\section{Methods}\label{methods}

\subsection{Sample}\label{sample}

Data for this study come from the third wave (2019) of the $PCSVDF^{Mulher}$ (Pesquisa de Condições Socioeconômicas e Violência Doméstica e Familiar contra a Mulher - Survey of Socioeconomic Conditions and Domestic and Family Violence against Women). The $PCSVDF^{Mulher}$ is an interdisciplinary effort to build empirical longitudinal evidence that enables the study of violence against women (including IPV), its causes and consequences for direct and indirect victims, the allocation of resources in the household, women and children's health; child development; and the interrelationships among them through an ethical and methodological sound approach \citep{Carvalho2018}.

The third wave of $PCSVDF^{Mulher}$ was approved by the Brazilian Ethics Committee (CONEP) under the number 20237519.0.0000.5054 and was carried out between 5 August 2019 and 15 December 2019 in the cities of Fortaleza / CE, Recife / PE, Salvador / BA, Goiânia / GO, Porto Alegre / RS, So Paulo / SP and Belém / PA.

According to \citet{carvalho_2023}, $PCSVDF^{Mulher}$ uses a complex, stratified, multistage probability cluster sampling design with unequal selection probabilities to select participants in wave I. In the first stage, a random sample of neighborhoods was selected. In the second stage, a random sample of census tracts was selected within these neighborhoods. In the third stage, households were selected within these census tracts. Finally, a woman aged 15 to 59 years was selected in each selected household to answer the questionnaire. In Wave II, $PCSVDF^{Mulher}$ interviewers attempted to reinterview the same women who had participated in Wave I. In cases where they were unable to reinterview the same woman, replacement by another woman within the same household or by another woman from a different household was allowed. In the latter case, the choice was made using the same protocol as in Wave I.

In wave III, the cities of Goiânia / GO, Porto Alegre / RS, So Paulo / SP and Belém / PA participated for the first time in $PCSVDF^{Mulher}$. Thus, the protocol used to select participants in these cities was the same protocol used in Wave I. For the cities of Fortaleza / CE, Recife / PE and Salvador / BA, which have been in $PCSVDF^{Mulher}$ since Wave I, the same protocol as in Wave II was used. The final sample of $PCSVDF^{Mulher}$ Wave III consisted of 8,447 women.

\subsection{Measures}\label{measures}

\subsubsection*{Dependent Variables}

The approach used in this paper requires two dependent variables: The first is victimization, and the second is reporting to the police. The $PCSVDF^{Mulher}$ survey assesses victimization using questions adapted from \citet{WHO2005} to assess the prevalence of domestic violence. Thus, for each type of violence (emotional, physical, or sexual), a series of acts were listed and asked, one by one, whether the interviewee had already suffered any of them at some point in her life. In case of an affirmative answer, we ask whether the act occurred in the last 12 months. Thus, the dichotomous variable \textit{violence12} was coded as 1 if the respondent suffered any type of violence in the last 12 months and 0 if the respondent did not. For the second variable, $PCSVDF^{Mulher}$ asked women who have been victims of violence in the last 12 months if ``In the last 12 months, have you ever gone to any of the following for help due to domestic violence?'' and listed a series of service networks for women in situations of violence. If the respondent answered that she sought help at Ordinary police stations, Military police, or Specialized Police Offices for Assistance to Women, then the dichotomous variable \textit{report\_police} was coded as 1, and 0 if she did not seek help at any of these locations.

\subsubsection*{Independent variables}

The independent variables in the current study are: \textit{Know Victim}, a dichotomous variable coded as 1 if the respondent knows any woman from her social network (i.e., acquaintance or woman that she is friendly with) who has been a victim of physical domestic violence in the past 12 months committed by a man. \textit{Mother Victim}, a dichotomous variable that takes the value 1 if, during the respondent's childhood, her mother was a victim of domestic violence. \textit{Police Perception} is a categorical variable that measures the respondents' perception about police patrol in the past 30 days in their neighborhood (0 = Never, 1 = Once/Twice, 2 = 3–5 times, 3 = 6–10 times, 4 = 11–20 times, 5 = More than 20 times). The variable \textit{Know DEAM} is a categorical variable that measures the respondent's knowledge about the Women's Specialized Police Station (DEAM) (= 1 if never heard of, = 2 if heard a little, 3 = if heard a lot). The variable \textit{Tolerate Violence} is a categorical variable that measures women's attitudes about gender relations. In the $PCSVDF^{Mulher}$ survey, the respondents were asked ``If necessary, do you believe that you should tolerate/accept physical violence in order to keep your family together?'' (1 = Agree, 2 = Neither agree nor disagree, 3 =  Disagree). To measure the impact of the type of violence on the probability of reporting, we also include the variables \textit{Vio\_Emo\_12} (=1 if the respondent suffered any act of emotional violence in the last 12 months), \textit{Vio\_Fis\_12} (=1 if the respondent suffered any act of physical violence in the last 12 months), and \textit{Vio\_Sex\_12} (=1 if the respondent suffered any act of sexual violence in the last 12 months). We also include partner characteristics, such as \textit{Partner Alcohol Abuse} (= 1 if drink every day or nearly every day or once or twice a week) and \textit{Partner Job} (= 1 if employed). Finally, we also include some respondent's demographic characteristics, such as \textit{Age} (in years), \textit{White} (= 1 if White or Asian), \textit{Education} (1 = No education, 2 = Incomplete or some fundamental school, 3 = Fundamental school, 4 = Incomplete or some high school, 5 = High school, 6 = Technical course, 7 = Incomplete or some University/College, 8 = University/College degree, 9 = At least some graduate program - treated as a continuous variable), \textit{Job} (= 1 if employed), \textit{Married} (= 1 if currently married), \textit{Children} (Number of children), \textit{Alchool Abuse} (= 1 if drink every day or nearly every day or once or twice a week), and \textit{City} (1 = Fortaleza/CE, 2 = Recife/PE, 3 = Salvador/BA, 4 = Goiânia/GO, 5 = Porto Alegre/RS, 6 = São Paulo/SP, 7 = Belém/PA)

\subsection{Analytic strategy}\label{analytic_strategy}

The analysis used in this study consisted of the following steps. First, we analyzed the prevalence of domestic violence disaggregated by type in the last 12 months. Then, we divided our sample into two groups (victims versus non-victims) and performed a chi-square test of independence and independent sample t-tests to check whether there were differences in the independent variables in the two groups. Third, we analyzed the reasons that victims of domestic violence used to report or did not report violence suffered to the police. Fourth, we again use a chi-square test of independence and independent sample t-tests to check whether there are differences in the independent variables now between the group of women victims of violence who reported the violence suffered and the group of women victims who did not report it. Finally, to understand the determinants of underreporting, we will use an approach similar to that used by \citet{MacDonald2001}, with the estimation of a bivariate probit regression model with sample selection. The choice of this model is due to the fact that victimization is highly correlated with socioeconomic factors that influence crime reporting behavior \citep{MacDonald2001}.

Therefore, first consider Eq. (\ref{eq_latent}), which relates the unobservable individual tendency to report with observable individual characteristics:

\begin{equation}\label{eq_latent}
r_{i}^{*} = \beta' x_{i} + \varepsilon_{i}
\end{equation}

\noindent Where $r_{*}^{i}$ is the latent variable that denotes the individual propensity to report a crime, $x_{i}$ is a vector of individual attributes that may contain socioeconomic characteristics of the interviewee, presence in the neighborhood of care services for women victims of violence, and the severity of the violence suffered, $\beta$ is the vector of parameters to estimate, and $\varepsilon_{i}$ is the normally distributed error term with zero mean and variance one, conditioned on $x$, that captures the unobserved determinants of the propensity to report. The latent variable $r_{i}^{*}$ determines the observable result of the decision to report or not, through Equation (\ref{eq_observada}).

\begin{equation}\label{eq_observada}
r_{i} = \begin{cases} 1, & r_{i}^{*} > 0 \\
0, & r_{i}^{*} \leq 0 \end{cases}
\end{equation}

Eq. (\ref{eq_latent}) can be estimated by a simple \textit{probit} or \textit{logit} model, with the results providing us with a direct measure of the impact of the explanatory variables on the probability of reporting the violence suffered. However, as suggested by \citep{MacDonald2001}, we can only estimate Eq. (\ref{eq_latent}) if we observe an episode of violence and if this is completely random, the estimation would not be a problem. However, being a victim of domestic violence or not is not a completely random event. There are individual characteristics of the woman and her partner that affect the probability of a woman being a victim of domestic violence (See \citet{Yakubovich2018}, \citet{Dias2020}, \citet{Tiruye2020}, \citet{Kebede2022}, \citet{CDC2021}, among others). Suppose then that the probability of being a victim of domestic violence is represented by a latent variable $b_{i}^{*}$, which is determined by:

\begin{equation}\label{eq_lante_vio}
b_{i}^{*} = \gamma' z_{i} + \mu_{i}
\end{equation}

\noindent where $z_{i}$ is a vector of socioeconomic variables of the woman and her partner, $\gamma$ is the vector of parameters to estimate, and $ \mu_{i}$ is the normally distributed error term with mean zero and variance one, conditioned on $z$. Like in Eq. (\ref{eq_latent}), the latent variable $b_{i}^{*}$ is related to the result observed of violence through Eq. (\ref{eq_observada_vio})

\begin{equation}\label{eq_observada_vio}
b_{i} = \begin{cases} 1, & b_{i}^{*} > 0 \\
0, & b_{i}^{*} \leq 0 \end{cases}
\end{equation}

\noindent So, note that we will only observe the result of reporting or not when there is violence, that is, when $b_{i} = 1 (b_{i}^{*} > 0)$.

The Separate Estimation of Eq. (\ref{eq_latent}) and Eq. (\ref{eq_lante_vio}) assumes that the covariance $\rho$ between $\varepsilon$ and $\mu$ is zero. However, if there are unobserved characteristics that influence both $r_{i}$ and $b_{i}$, then $\varepsilon$ and $\mu$ will be correlated, that is, $\rho \neq 0$ and the estimates obtained by a univariate \textit{probit} (or \textit{logit}) model will be biased. Thus, to avoid this problem, we can estimate a bivariate \textit{probit} from the joint results of Eq. (\ref{eq_latent}) and Eq. (\ref{eq_lante_vio}), where the joint probability of being a victim of domestic violence and reporting the incident is given by:

\begin{equation}\label{eq_probit_biv}
Pr(r_{i} = 1, b_{i} = 1) = \Phi(\beta'x_{i}, \gamma'z_{i}, \rho)
\end{equation}

\noindent where $\Phi$ is the cumulative distribution function of a bivariate normal distribution. Empirically, this model will be identified if there is an exclusion restriction; that is, we need a variable in $z_{i}$ that influences the probability of being a victim of domestic violence but that does not impact the propensity to report the incident, that is, not in $x_{i}$ \citep{Maddala2010}.

\section{Results}\label{results}

\subsection*{Prevalence of IPV}

As stated previously, the main objective of this paper is to understand the under-reporting of IPV in Brazil. Therefore, we first need to understand the prevalence of IPV. The Table (\ref{tab:prevalences}) presents the prevalence of IPV disaggregated by type of violence in the last 12 months in our sample.

\begin{table}[H]
\caption{\label{tab:prevalences} Prevalence of IPV disaggregated by type of violence in the last 12 months}
\centering
\begin{tabu} to \linewidth {>{\raggedright\arraybackslash}p{7cm}>{\centering}X>{\centering}p{3cm}>{\centering}X>{\centering}X}
\toprule
Type of Violence & \% & CI & n (Unw) & N (Unw) \\
\midrule
Emotional & 13.77 & 12.35 - 15.33 & 528 & 3618\\
Physical & 4.92 & 4.1 - 5.89 & 208 & 3576\\
Sexual & 2.77 & 2.16 - 3.56 & 108 & 3583\\
Any type & 15.55 & 14.05 - 17.19 & 596 & 3621\\
\bottomrule
\multicolumn{5}{l}{\rule{0pt}{1em}\textit{Source: } Elaborated by the authors}\\
\end{tabu}
\end{table}

As we can see, 15.55\% (n = 596) of the women in our sample were victims of IPV in the last 12 months. By disaggregating by type of violence, we can see that the most frequent type of violence was emotional, with 13.77\% (n = 528) of women having suffered this type of violence in the last 12 months. Physical violence appears next, with 4.92\% (n = 208), and then sexual violence, with 2.77\% (n = 108).

Table (\ref{tab:violence_factors}) shows the difference in the composition of the group of women who were victims and those who were not victims of IPV. The results indicate a statistically significant difference in some variables between the two groups. The \textit{Violence Group} is slightly younger (Dif = -1.84, p = 0.008), with a slightly lower level of education (Dif = -0.24, p = 0.0737), a lower percentage of married women (Dif = -6.53, p = 0.0255), and a higher average number of children (Dif = 0.17, p = 0.0311) than the \textit{Non-Violence Group}.

\begin{table}[H]
\caption{\label{tab:violence_factors} Descriptive statistics for factors associated with violence versus non violence}
\centering
\begin{tabu} to \linewidth {>{\raggedright\arraybackslash}p{3cm}>{\raggedright\arraybackslash}p{4.5cm}>{\centering}X>{\centering}X>{\centering}X>{\centering}X>{\centering}X>{\centering}X}
\toprule
\multicolumn{1}{c}{Variable} & \multicolumn{1}{c}{Category} & \multicolumn{2}{c}{Victim} & \multicolumn{2}{c}{Non Victim} & \multicolumn{2}{c}{$t$-test} \\
\cmidrule(l{3pt}r{3pt}){3-4} \cmidrule(l{3pt}r{3pt}){5-6} \cmidrule(l{3pt}r{3pt}){7-8}
 & & \% & se & \% & se & Dif & $p$-val\\
\midrule
Age & In years & 34.57 & 0.66 & 36.4 & 0.3 & -1.84 & 0.008\\
White & 1 = White and Asian & 29.25 & 2.85 & 32.91 & 2.21 & -3.66 & 0.178\\
Education & 1 = No education ... 9 = Postgraduate & 5.44 & 0.13 & 5.68 & 0.07 & -0.24 & 0.0737\\
Job & 1 = Employed & 64.45 & 2.61 & 60.62 & 1.37 & 3.83 & 0.1629\\
Married & 1 = Yes & 32.01 & 2.75 & 38.54 & 1.49 & -6.53 & 0.0255\\
Children & Number of Children & 1.48 & 0.07 & 1.31 & 0.04 & 0.17 & 0.0311\\
Alcohol Abuse & 1 = Yes & 25.01 & 2.64 & 17.55 & 1.43 & 7.47 & 0.0025\\
Know Victim  & 1 = Yes & 37.64 & 2.61 & 25.3 & 1.28 & 12.35 & 0\\
Mother Victim  & 1 = Yes & 41.63 & 3.34 & 25.91 & 1.15 & 15.73 & 0\\
Police Perception & 0 = Never ... 5 = +20 times & 2.29 & 0.11 & 2.27 & 0.06 & 0.02 & 0.8465\\
Partner Alchool Abuse & 1 = Yes & 45.23 & 2.62 & 35.84 & 1.65 & 9.39 & 0.0008\\
Partner Job & 1 = Employed & 76.99 & 1.99 & 77.58 & 1.13 & -0.59 & 0.794\\
\midrule
\multicolumn{1}{c}{Variable} & \multicolumn{1}{c}{Category} & \multicolumn{2}{c}{Victim} & \multicolumn{2}{c}{Non Victim} & \multicolumn{2}{c}{$\chi^{2}$-test} \\
\cmidrule(l{3pt}r{3pt}){3-4} \cmidrule(l{3pt}r{3pt}){5-6} \cmidrule(l{3pt}r{3pt}){7-8}
& & \% & N(Unw) & \% & N(Unw) & F & $p$-val\\
\midrule
Know DEAM    & Never & 6.37 & 51 & 7.35 & 267 & 0.3721 & 0.6896\\
             & A few & 31.9 & 188 & 32.14 & 995\\
             & Many times & 61.74 & 352 & 60.5 & 1751\\
Tolerate Violence & Agree & 10.83 & 60 & 3.54 & 121 & 8.5783 & 0.0002\\
                  & Neither agree nor disagree & 1.88 & 13 & 0.96 & 32\\
                  & Disagree & 87.29 & 518 & 95.5 & 2848\\
City         & Fortaleza/CE & 16.74 & 100 & 18.6 & 582 & 1.3179 & 0.2488\\
             & Recife/PE & 12.13 & 82 & 13.48 & 391\\
             & Salvador/BA & 18.33 & 102 & 20.22 & 619\\
             & Goiânia/GO & 14.27 & 80 & 10.67 & 328\\
             & Porto Alegre/RS & 11.54 & 77 & 12.68 & 355\\
             & São Paulo/SP & 14.7 & 79 & 14.99 & 446\\
             & Belém/PA & 12.28 & 76 & 9.36 & 304\\
\bottomrule
\multicolumn{8}{l}{\rule{0pt}{1em}\textit{Source: } Elaborated by the authors}\\
\end{tabu}
\end{table}

Concerning alcohol abuse, the \textit{Violence Group} has a higher percentage of both women who abuse alcohol (Dif = 7.47, p = 0.0025) and their partners or ex-partners (Dif = 9.39, p = 0.0008).

Among the \textit{Violence Group}, there is a higher percentage of women who know other women in their social circle who were also victims of physical violence than in the \textit{Non-Violence Group} (Dif = 12.35, p = 0.0000). Furthermore, the percentage of women who knew that their mothers suffered IPV during their childhood is also higher in the \textit{Violence Group} (Dif = 15.73, p = 0.0000). 

Finally, we can see that among the \textit{Violence Group}, there is a higher percentage of women who believe that they should tolerate violence to keep the family together. In this group, 10. 83\% of the women agreed, compared to only 3. 54\% in the \textit{ non-violence group} (F = 8.5783, p = 0.0002). 

On the other hand, the variables \textit{White}, \textit{Job}, \textit{Police Perception}, \textit{Partner Job}, \textit{Know DEAM}, and \textit{City} did not show statistically significant differences between the two groups.

\subsection*{Reporting the IPV to the police}

As shown in Table (\ref{tab:prevalences}), 596 women in our sample were victims of IPV in the last 12 months. However, only 71 (11.91\%) reported the incident to the police, while 376 (63.09\%) did not report it, and 149 (25\%) did not want to answer. This difference between the number of incidents that occurred and the number of incidents reported is what experts call the ``dark figure'' \citep{skogan_contacts_1994}. Tables (\ref{tab:reason_call_police}) and (\ref{tab:reason_not_call_police}) below show, respectively, the main reasons that IPV victims in our sample reported or did not report the incident to the police.

\begin{table}[H]
\caption{\label{tab:reason_call_police} Reasons for Calling the Police [N = 71 (11.91\%)]}
\centering
\begin{tabu} to \linewidth {>{\raggedright\arraybackslash}p{8cm}>{\centering}X>{\centering}X}
\toprule
Reason & \% & N (Unw) \\
\midrule
Encouraged by friends/family & 28.12 & 22\\
Could not endure more & 44.91 & 36\\
Badly injured & 7.27 & 3\\
He threatened or tried to kill you & 16.29 & 11\\
He threatened or hit children & 7.16 & 6\\
Saw that children were suffering & 8.62 & 6\\
Thrown out of the home & 1.48 & 2\\
Afraid you would kill him & 0.52 & 1\\
Afraid he would kill you & 11.41 & 10\\
Other & 21.49 & 11\\
\bottomrule
\multicolumn{3}{l}{\rule{0pt}{1em}\textit{Source: } Elaborated by the authors}\\
\end{tabu}
\end{table}

The reason most cited by women who reported IPV to the police was ``Could not endure more'' with 44.91\% (n = 36), followed by ``Encouraged by friends/family'' (28.12\%, n = 22) and ``He threatened or tried to kill you'' (16.29\%, n = 11). Furthermore, 11. 41\% (n = 10) of the women in our sample stated that they reported it because they were ``afraid that he would kill you''.

Regarding the reasons for not reporting IPV to the police, the three most cited reasons were the following. Firstly, many victims did not consider the incident serious enough to warrant the involvement of the police (approximately 67\%). Secondly, some victims were afraid of the consequences of reporting violence, such as further violence or threats (approximately 5\%). Third, they were either too embarrassed, ashamed or afraid to report the incident, fearing that they would not be believed or blamed for the violence (approximately 3\%).

\begin{table}[H]
\caption{\label{tab:reason_not_call_police} Most Important Reasons for Not Calling the Police [(N = 376 - 63.09\%)]}
\centering
\begin{tabu} to \linewidth {>{\raggedright\arraybackslash}p{8cm}>{\centering}X>{\centering}X>{\centering}X>{\centering}X>{\centering}X>{\centering}X}
\toprule
\multicolumn{1}{c}{Reason} & \multicolumn{2}{c}{Police Patrol} & \multicolumn{2}{c}{Police Station} & \multicolumn{2}{c}{DEAM}\\
\cmidrule(l{3pt}r{3pt}){2-3} \cmidrule(l{3pt}r{3pt}){4-5} \cmidrule(l{3pt}r{3pt}){6-7}
 & \% & N (Unw) & \% & N (Unw) & \% & N (Unw) \\
\midrule
I did not consider what happened to me a serious incident & 66.89 & 237 & 67.28 & 241 & 67.14 & 241\\
Embarrassed/Ashamed/Afraid would not be believed or would be blamed & 3.29 & 21 & 2.93 & 17 & 3.35 & 19 \\
Fear or threats/Consequences/More violence & 5.28 & 18 & 5.06 & 19 & 5.32 & 19\\
Believed it would not help & 3.92 & 15 & 4.22 & 15 & 4.03 & 14\\
Know other women that were not helped & 0.45 & 4 & 0.94 & 6 & 0.1 & 1\\
Afraid he would be arrested & 3.23 & 14 & 2.78 & 12 & 1.41 & 7\\
This service does not exist in the city/community & - & - & 0.39 & 2 & 1.05 & 4\\
Don't know/Didn't answer & 16.94 & 58 & 16.38 & 55 & 17.58 & 62\\
\bottomrule
\multicolumn{7}{l}{\rule{0pt}{1em}\textit{Source: } Elaborated by the authors}\\
\end{tabu}
\end{table}

Now, we will turn our attention to analyzing the difference in the composition of the group of women who reported IPV to the police and those who did not.

Table (\ref{tab:report_factors}) highlights significant differences between the two groups. In the \textit{Reporting group}, we can see a higher percentage of white women than in the \textit{Non-Reporting group} (Dif = 15.79, p = 0.0561), just as there is a lower percentage of married women (Dif = -22.86, p = 0.0008). Concerning the variable \textit{Know DEAM}, we found that in the \textit{Reporting group}, there is a higher percentage of women who heard \textit{A lot} about DEAM than in the \textit{Non-Reporting group} (F = 3.7116, p = 0.0255). In addition, as expected, we found that women who believe that they should tolerate domestic violence to keep their family together are present in a higher percentage in the \textit{ non-reporting group} than in the \textit{ report group} (F = 7.7715, p = 0.0005). Finally, we also found that in the groups of women who reported IPV to the police, there was a higher percentage of women who suffered physical violence (Dif = 26.01, p = 0.0081).

\begin{table}[H]
\caption{\label{tab:report_factors} Descriptive statistics for factors associated with police notification for the victims who reported the IPV to the police}
\centering
\begin{tabu} to \linewidth {>{\raggedright\arraybackslash}p{3cm}>{\raggedright\arraybackslash}p{4.5cm}>{\centering}X>{\centering}X>{\centering}X>{\centering}X>{\centering}X>{\centering}X}
\toprule
\multicolumn{1}{c}{Variable} & \multicolumn{1}{c}{Category} & \multicolumn{2}{c}{Reported} & \multicolumn{2}{c}{Not Reported} & \multicolumn{2}{c}{$t$-test} \\
\cmidrule(l{3pt}r{3pt}){3-4} \cmidrule(l{3pt}r{3pt}){5-6} \cmidrule(l{3pt}r{3pt}){7-8}
 & & \% & se & \% & se & Dif & $p$-val\\
\midrule
Age & In years & 35.69 & 2.06 & 34.36 & 0.85 & 1.33 & 0.5462\\
White & 1 = White and Asian & 41.89 & 7.36 & 26.1 & 3.52 & 15.79 & 0.0561\\
Education & 1 = No education ... 9 = Postgraduate & 5.39 & 0.4 & 5.53 & 0.16 & -0.14 & 0.7478\\
Job & 1 = Employed & 75.74 & 6.09 & 65.92 & 3.33 & 9.82 & 0.1651\\
Married & 1 = Yes & 14.97 & 5.5 & 37.83 & 3.7 & -22.86 & 0.0008\\
Children & Number of Children & 1.59 & 0.2 & 1.43 & 0.1 & 0.16 & 0.497\\
Alcohol Abuse & 1 = Yes & 28.39 & 6.84 & 24.01 & 2.97 & 4.38 & 0.5416\\
Know Victim  & 1 = Yes & 48.13 & 7.26 & 37.35 & 3.14 & 10.78 & 0.1669\\
Mother Victim  & 1 = Yes & 34.96 & 8.63 & 43.57 & 3.63 & -8.61 & 0.3401\\
Police Perception & 0 = Never ... 5 = +20 times & 2.22 & 0.38 & 2.23 & 0.13 & -0.02 & 0.9623\\
Partner Alchool Abuse & 1 = Yes & 51.77 & 7.29 & 46.19 & 3.41 & 5.59 & 0.4947\\
Partner Job & 1 = Employed & 71.17 & 6.34 & 79.51 & 2.38 & -8.34 & 0.2342\\
\midrule
\multicolumn{1}{c}{Variable} & \multicolumn{1}{c}{Category} & \multicolumn{2}{c}{Reported} & \multicolumn{2}{c}{Not Reported} & \multicolumn{2}{c}{$\chi^{2}$-test} \\
\cmidrule(l{3pt}r{3pt}){3-4} \cmidrule(l{3pt}r{3pt}){5-6} \cmidrule(l{3pt}r{3pt}){7-8}
& & \% & N(Unw) & \% & N(Unw) & F & $p$-val\\
\midrule
Know DEAM    & Never & 1.71 & 3 & 7.03 & 33 & 3.7116 & 0.0255\\
             & A little & 22.89 & 19 & 31.86 & 118\\
             & A lot & 75.4 & 48 & 61.11 & 223\\
Tolerate Violence & Agree & 1.15 & 2 & 14.3 & 51 & 7.7715 & 0.0005\\
                  & Neither agree nor disagree & 1.31 & 1 & 2.44 & 11\\
                  & Disagree & 97.53 & 68 & 83.26 & 313\\
City         & Fortaleza/CE & 13.91 & 14 & 17.49 & 66 & 0.3849 & 0.8885\\
             & Recife/PE & 15.66 & 12 & 12.5 & 52\\
             & Salvador/BA & 19.13 & 13 & 22.07 & 79\\
             & Goiânia/GO & 11.29 & 10 & 13.77 & 47\\
             & Porto Alegre/RS & 21.61 & 9 & 8.6 & 39\\
             & São Paulo/SP & 11.07 & 8 & 14.78 & 50\\
             & Belém/PA & 7.33 & 5 & 10.78 & 43\\
\midrule
\multicolumn{2}{l}{Type of Violence} & \multicolumn{2}{c}{Reported} & \multicolumn{2}{c}{Not Reported} & \multicolumn{2}{c}{$t$-test} \\
\cmidrule(l{3pt}r{3pt}){3-4} \cmidrule(l{3pt}r{3pt}){5-6} \cmidrule(l{3pt}r{3pt}){7-8}
 & & \% & N(Unw) & \% & N(Unw) & Dif & $p$-val\\
\midrule
\multicolumn{2}{l}{Emotional Violence}                          & 92.00 & 64 & 88.97 & 329 & 3.03 & 0.4988\\
\multicolumn{2}{l}{Physical Violence}                           & 54.36 & 36 & 28.36 & 116 & 26.01 & 0.0081\\
\multicolumn{2}{l}{Sexual Violence}                             & 18.24 & 16 & 17.68 & 68 & 0.55 & 0.9291\\
\bottomrule
\multicolumn{8}{l}{\rule{0pt}{1em}\textit{Source: } Elaborated by the authors}\\
\end{tabu}
\end{table}

\subsection*{Econometric Model}

Before presenting the results of the bivariate probit with sample selection model, it is worth stressing that we utilized an adaptive Lasso for variable selection, as described in \citet{Ogundimu2022}. We utilized the R package \textit{HeckmanSelect} \citep{Ogundimu2024}. The following table (\ref{tab:model_results}) shows the result of the model with the variables chosen by adaptive Lasso for the selection and outcome equations. The final model was estimated using the R package \textit{SampleSelection} \citep{Henningsen2008} and employing the simulated annealing algorithm.

\begin{table}[H]
\caption{\label{tab:model_results} Results of the bivariate probit with sample selection model}
\centering
\begin{tabu} to \linewidth {>{\raggedright\arraybackslash}p{6cm}>{\centering}X>{\centering}X}
\toprule
 & \multicolumn{2}{c}{\textit{Dependent Variable}} \\ 
\cline{2-3} 
\\[-1.8ex]
 & \multicolumn{1}{c}{Violence} & \multicolumn{1}{c}{Report to police} \\ 
\\[-2.8ex]
\midrule
  Intercept & -0.3550 (0.2718) & 0.4738 (0.4664) \\ 
  Age & $-0.0136^{***}$ (0.0030) & $0.0150^{***}$ (0.0052) \\ 
  White & - & $0.2741^{**}$ (0.1229) \\ 
  Education & - & -0.0266 (0.0284) \\ 
  Job & $0.1037^{*}$ (0.0604) & - \\ 
  Married & -0.0886 (0.0670) & $-0.2406^{**}$ (0.1220) \\ 
  Children & $0.0908^{***}$ (0.0230) & -0.0321 (0.0410) \\ 
  Know Victim & - & 0.0793 (0.0994) \\ 
  Mother Victim & $0.3488^{***}$ (0.0650) & $-0.3209^{***}$ (0.1060) \\ 
  Police Perception & - & 0.0063 (0.0268) \\ 
  Partner Alchool Abuse & $0.2661^{***}$ (0.0634) & $-0.2202^{**}$ (0.1041) \\ 
  Partner Job & -0.0471 (0.0684) & - \\ 
  Emotional Violence & - & 0.1673 (0.1625) \\ 
  Physical Violence & - & $0.2459^{**}$ (0.1085) \\ 
  Sexual Violence & - & 0.1118 (0.1373) \\ 
  \textbf{Know DEAM} & \\
  \hspace{0.25cm}Never (ref) & - & - \\
  \hspace{0.25cm}A little & - & 0.1166 (0.1951) \\ 
  \hspace{0.25cm}A lot & - & 0.2535 (0.1876) \\ 
  \textbf{Tolerate Violence} & \\
  \hspace{0.25cm}Neither agree nor disagree (ref) & - & - \\
  \hspace{0.25cm}Agree & -0.0597 (0.2583) & -0.0991 (0.3598) \\ 
  \hspace{0.25cm}Disagree & $-0.7435^{***}$ (0.2351) & $0.5960^{*}$ (0.3320) \\ 
\midrule
Observations & \multicolumn{1}{c}{3,049} & \multicolumn{1}{c}{374} \\ 
Log Likelihood & \multicolumn{2}{c}{-1442.785} \\ 
$\rho$ & \multicolumn{2}{c}{$-0.7901^{***}$ (0.1031)} \\ 
\bottomrule
\multicolumn{3}{l}{\rule{0pt}{1em}\textit{Source:} Elaborated by the authors}\\
\multicolumn{3}{l}{\rule{0pt}{1em}\textit{Note:} $^{*}$p$<$0.1; $^{**}$p$<$0.05; $^{***}$p$<$0.01}
\end{tabu}
\end{table}

For the selection equation, except \textit{Married} and \textit{Partner job}, all other variables are statistically significant. We found that younger women are more likely to suffer from IPV, as well as women who have a job. Having children also increases the probability of women suffering from IPV. As expected, having a partner who abuses alcohol increases the probability of the woman suffering IPV, and not tolerating domestic violence decreases the probability of the woman suffering IPV. Finally, having a mother who was a victim of domestic violence also increases the likelihood of the woman suffering IPV.

For the outcome equation, as expected, victims of physical IPV and women who do not tolerate domestic violence are more likely to report the incident to the police. However, having a partner who abuses alcohol and being married decreases the probability of the woman reporting the violence suffered to the police. Regarding sociodemographic characteristics, white and older women are more likely to report IPV. Finally, having a mother who suffered domestic violence decreases the likelihood of the woman reporting IPV to the police. All other variables do not seem to play an important role in a woman's decision to report the IPV to the police.

\section{Discussion and Conclusions}\label{discussion}

The underreporting of cases of domestic violence in Brazil is still unknown, with few studies seeking to estimate this number. \citet{Senado2024} estimates the number of women who were victims of physical and/or sexual violence and did not report it to the authorities at 61\%. \citet{Vasconcelos2023} estimate that 75.9\% of victims of physical violence do not report the violence they suffered. This number increases even more for cases of sexual (89.4\%) and psychological (98.5\%) violence. Therefore, to shed light on this topic, this study sought to estimate the underreporting of IPV in Brazil and identify the factors associated with the decision to report or not the violence suffered to the police.

We found that only 11. 91\% of the IPV victims report the case to the police. Of the women who reported it, the two main reasons for asking the police for help were because they could no longer support the violent situation they were experiencing and the support they received from family and friends. Thus, we can highlight two important points from these results: i) The act of reporting the violence suffered seems to have a component of resistance, where the woman does not report the violence suffered, believing that the perpetrator will change his behavior, resisting reporting until a breaking point, where she can no longer bear the aggression. For example, \citet{podana_reporting_2010} showed that the history of incidents is a key variable in explaining the decision to report IPV to the police; ii) Having a support network is essential for women who are victims of violence to break the cycle of violence. \citet{jong_desistindo_2008} identified that many IPV victims who filed a complaint against the aggressor resorted to the help of friends and family to support their complaint.

Concerning IPV victims who did not report, we found that the vast majority did not report it because they considered that the violence suffered was not serious. This finding is consistent with \citet{felson_reasons_2002}, which also found that the main reason for not calling the police was because the victim believed that the incident was trivial. \citet{akers_police_2009} also found that women who called the police were more likely to have experienced severe forms of violence. Another important finding is that the second most cited reason not to report to police is fear of more violence. \citet{felson_reasons_2002} also found that fear of reprisal is one important key to not reporting violence to the police. 

In Brazil, the Maria da Penha law \citep{lei2006cria} created mechanisms to protect women victims of violence by implementing the so-called protective measures. Among the measures imposed on the aggressor are the suspension or restriction of carrying weapons and the prohibition of approaching and contacting the victim. Despite the progress made, unfortunately, these measures were not enough. In 2023, 20\% of the victims of femicide in Porto Alegre / RS had protective measures \citep{GZH2023}. In Fortaleza / CE, 9. 5\% of the victims of attempted femicide had protective measures \citep{Quezado2023}. These results show that the Brazilian government needs to improve the application of the Maria da Penha law to ensure that victims of IPV feel safe to report the violence they suffered.

In order to answer our question about the factors associated with the report decision, we estimated a bivariate probit with sample selection. In this model, the first stage, that is, the selection equation, estimates the factors associated with the probability of a woman being a victim of IPV. We found that older women are less likely to be victims of IPV. In a systematic review, \citet{Capaldi2012} found relatively consistent results that age is protective against IPV in adulthood. \citet{Yakubovich2018} also found that being older is a protective factor against women experiencing IPV.

An interesting result we found is that employed women are more likely to become victims of IPV. This result is opposite to the one found by \citet{Aizer2010} and \citet{Anderberg2015}. \citet{Aizer2010} found that decreases in the wage gap reduce violence against women, and \citet{Anderberg2015} found that an increase in the female unemployment rate causes an increase in the incidence of physical abuse against women. On the other hand, our results are similar to those found by \citet{Bhalotra2019} who, using representative data from 31 developing countries, found that a decrease in female unemployment rates is associated with an increase in the probability of victimization.

We also found that women whose mothers were victims of domestic violence are more likely to be victimized. This result was expected, as studies such as \citet{whitfield_violent_2003}, \citet{pollak_intergenerational_2004}, and \citet{ludermir_previous_2017} show that children who grow up exposed to IPV are more likely to become perpetrators (boys) or victims (girls) when they reach adulthood compared to children who were not exposed to IPV. We also found a positive association between having children and being a victim of IPV. This result is worrying, as it intensifies the intergenerational cycle of violence.

Another important finding is that women whose partners/ex-partners have problems with alcoholic beverages are more likely to be victims of domestic violence. It should be noted that the literature indicates that the use of alcohol by one or both partners contributes to increasing the risk and severity of IPV \citep{wilson_alcohol_2014, ally_intimate_2016, sontate_alcohol_2021}.

Concerning the second stage, i.e., the outcome equation, we estimated the factors associated with the probability of reporting IPV to the police, given that the woman is a victim of IPV. Our results show that although age is a protective factor for the risk of IPV, once it has occurred, it becomes a relevant factor for the decision to report it to the police, i.e. older women tend to report more of the violence suffered. This result corroborates what was found by \citet{akers_police_2009}.

Regarding race, we found that white women are more likely to report IPV than black and brown women. This result is the opposite, as found by \citet{Holliday2019} and \citet{Powers2022}. Using data for the US, both found that black women are more likely to report IPV to the police than white women.

As in \citet{akers_police_2009}, we also found that married women are less likely to report IPV. \citet{akers_police_2009} states that married women have stronger emotional and financial ties to the relationship, which leads them to decide not to report the aggressor.

As we found in age, having a mother who was a victim of domestic violence has the opposite effect in being a victim of IPV and reporting IPV. We found that women whose mothers were victims of IPV are less likely to report IPV to the police. This result is in line with the hypothesis \citet{pollak_intergenerational_2004} that women who witness domestic violence as children are more likely to stay with an abusive spouse. A similar result, but with a focus on child abuse, was found by \citet{kim_factors_2022}. They found that women with experience of child abuse were less willing to report IPV to the police.

Having a partner who has problems with alcoholic beverages not only increases the likelihood of the woman becoming a victim of IPV but also decreases the probability that the woman reports IPV. A possible explanation is that, once alcohol consumption is associated with the risk and severity of IPV \citep{wilson_alcohol_2014, ally_intimate_2016, sontate_alcohol_2021}, victims of IPV are afraid of the retaliation that may occur because they have reported the aggressor to the police, fearing their physical integrity and that of their families.

Physical violence, as opposed to emotional and sexual violence, proved to be an important factor in explaining the decision to report the IPV suffered. Women who have suffered physical violence are more likely to report the aggressor to the police. These results corroborate the hypothesis that the act of reporting IPV is related to the severity of violence \citep{kim_factors_2022}. However, an addendum needs to be made here. Although sexual violence can cause very serious psychological damage, many women cannot perceive this type of violence within a relationship, and when they do, they are unable to report it, either out of shame or because they think the authorities will not believe them.

Finally, as expected, we found that women who do not tolerate violence are more likely to report IPV. This result is in line with \citet{Shakya2022}, which found that women who have reported IPV are more likely to report attitudes supporting IPV.

Considering our results, we can suggest two channels for action by authorities and public managers to reduce the number of unreported IPV cases. The first would be through educational actions so that women victims of IPV can identify that the acts they suffer are subject to reporting and punishment, given that approximately 67\% of the victims of IPV in our sample did not report the case to the police because they considered the incident not serious enough. In addition, we also found that married women and women whose mothers were victims of violence are less likely to report the violence they suffered, which reinforces the need for educational actions that show these women that acts of violence within relationships are not normal and should not be tolerated. The second channel of action is through actions that guarantee the safety of the victim and their relatives, since the second most cited reason for not reporting IPV to the police was fear of further violence, violence that the partner's alcohol consumption can aggravate. Thus, improving the granting and enforcement of protective measures established under the Maria da Penha Law could encourage victims of IPV to report the crime, reducing underreporting.

This article aimed to contribute to the literature on under-reporting of IPV in Brazil. We believe that we achieved our goals and contributed to the literature on this important topic for society. Despite that, it is critical to acknowledge its limitations. A significant limitation is the geographic scope of the data collection for the PCSVDFMulher sample. Since the data were collected only in 7 capitals, it was not possible to capture all the differences that exist between the different regions of Brazil. Therefore, it is not possible to extrapolate the results obtained in this study to the entire country, especially to medium and small cities in the interior of the country, which often do not have specialized services of the network of care for women in situations of violence, which discourages them from reporting the IPV they have suffered. Another limitation is that, due to the use of only the third wave of the PCSVDFMulher, we cannot draw any causal implication from our results. However, this is not our intention. Our objective was only to analyze the under-reporting of IPV cases and identify the main factors associated with it. For future studies, we can explore the longitudinal capacity of the PCSVDFMulher and try to establish causal relationships between the independent and dependent variables, in addition to analyzing the behavior of the IPV under-reporting rate over time.

\section*{Declaration of Conflicting Interests}

\noindent The author(s) declared that they have no potential conflicts of interest with respect to the research, authorship, and/or publication of this article.

\section*{Funding}

\noindent This study was financed in part by the Coordenação de Aperfeiçoamento de Pessoal de Nível Superior - Brasil (CAPES) - Finance Code 001.

\section*{Ethical Approval}

\noindent This article uses data from the third wave of the $PCSVDF^{Mulher}$ survey, which was approved by the Brazilian Ethical Committee (CONEP) under the number 20237519.0.0000.5054.

\section*{Informed Consent}

\noindent Informed consent was obtained from all participants of the $PCSVDF^{Mulher}$ survey. For participants under 18 years of age, informed written parental consent was obtained.

\section*{Data Availability}

\noindent The datasets generated during and/or analyzed during the current study are available from the corresponding author on reasonable request.

\bibliographystyle{apalike}
\bibliography{Dark_figure.bib}
\end{document}